\documentclass[prd,aps,preprint,tightenlines]{revtex4}
\usepackage{epsfig}
\usepackage{amsmath}
\begin{document}
\title{Validity Of The Vafa-Witten Proof \\
On Absence Of Spontaneous Parity Breaking In QCD}
\author{Xiangdong Ji}
\email{xji@physics.umd.edu}
\affiliation{Department of Physics,
University of Maryland,
College Park, Maryland 20742 }
\date{\today}          
\begin{abstract}

The vacuum energy response of a quantum field theory is studied as a 
function of complex external fields coupled to the symmetry-breaking
observables of the theory. It is found that the energy density is 
insensitive to spontaneous symmetry breaking effects along a
special direction in the complex plane. In particular, in 
the case of vector-like gauge theories such as QCD, 
the change in energy generated by coupling to pure-gluonic 
parity-odd operators is always a growing function of 
the external fields along the real axis, 
independent of the mode in which symmetry is realized.
In light of this, although there is little doubt 
that parity is not spontaneously broken in massive QCD, 
the validity of the Vafa and Witten proof is, however, 
questionable.

\vspace{10cm}
\end{abstract}
\maketitle
\newcommand{\be}{\begin{equation}}
\newcommand{\ee}{\end{equation}}
\newcommand{\ben}{\[}
\newcommand{\een}{\]}
\newcommand{\beqn}{\begin{eqnarray}}
\newcommand{\eeqn}{\end{eqnarray}}
\newcommand{\Tr}{{\rm Tr} }

Sometime ago Vafa and Witten made a proof that no spontaneous breaking 
of parity can occur to vector-like gauge theories such as
quantum chromodynamics (QCD) \cite{vafawitten}.  The argument is simple 
and elegant: Consider an external field $\lambda$ coupled to a
parity-odd observable $X$ constructed entirely out of gauge potentials. 
After applying hermiticity and Wick rotation, the extra term
in the field-theoretic action contributes to the Feynman path integral 
as a multiplicative, pure phase factor.  With a positive-definite
integrand as in vector-like gauge theories, the phase factor 
can only lower the value of the generating functional $Z$. Hence the vacuum
energy of the system must increase with the external coupling, 
and consequently spontaneous parity breaking can never happen.

Although there is little doubt that parity is not spontaneously 
broken in the real-world QCD, the Vafa-Witten proof has stimulated 
much discussion in the literature. Aoki found a parity-breaking
phase in lattice QCD with Wilson fermions which satisfies all
conditions of the Vafa-Witten proof \cite{aoki}. 
Sharpe and Singleton argued that the observables
constructed out of quark fields do have different properties 
from those of pure gluonic operators, and contrary to a comment made by Vafa
and Witten \cite{sharpe}. The question arises then why 
all the gluonic observables have zero expectation values in the
symmetry breaking phase. Along a different line of investigation, 
Azcoiti and Galante found that the vacuum energy density in the
presence of the external symmetry breaking source simply does not 
exist.  Therefore, the very starting point of the Vafa-Witten proof 
is in doubt \cite{azcoiti}. In a recent publication, Cohen \cite{cohen} showed 
that the Vafa-Witten proof does not apply to finite
temperature QCD systems because there exist observables
whose couplings to external fields do not contribute to the path-integral
through just pure phase factors. In fact, parity-breaking effects in
heavy-ion collisions have been discussed recently by Kharzeev,
Pisarski, and Tytgat \cite{kharzeev}.

In this paper, we show that Vafa-Witten's proof 
overlooked a special property of the vacuum energy response: 
this response is insensitive to spontaneous symmetry breaking effects 
along a special direction in the plane of complex probing fields. This 
direction happens to be the real axis in the case of coupling to 
parity-odd pure-gluonic observables. As such, a spontaneous parity breaking 
would have no effect upon the perturbed vacuum energy density. 
Symmetry-breaking effects, if there are, emerge only if the probing 
fields are analytically continued off the real axis. However, 
Vafa and Witten's argument no longer applies there. In short,
it may well be that the parity is conserved in the real world
QCD, however the Vafa-Witten proof is not satisfactory.

To begin, let us consider a Minkowski field theory with lagrangian 
density ${\cal L}$ which is invariant under a symmetry group G.
Suppose that in the vacuum the symmetry is spontaneously 
broken down to a subgroup H.  Let a constant field $\lambda$ couple 
to a Lorentz-invariant observable $X$ which is a singlet 
under H but not under G.  The perturbed lagrangian density reads,
\begin{equation}
          {\cal L}(\lambda) = {\cal L} + \lambda X \ .  
\end{equation}
The vacuum energy density $E(\lambda)$ of the perturbed system can be 
calculated as a Feynman path integral,
\begin{equation}
      \exp\left(-iE(\lambda)V_4\right) = \int [D\phi] 
\exp\left(i\int d^4x({\cal L}+\lambda X)\right)\ , 
\end{equation}
where $V_4$ is the four-volume and the limit $V_4\rightarrow \infty$ 
is implicitly understood. We are interested in the behavior of
$E(\lambda)$ near the origin of the complex $\lambda$ plane. 
We will assume that measure is positive-definite although most of our
results are independent of this assumption.

It is instructive to consider at first the vacuum energy response 
when there is no spontaneous symmetry breaking.  Then the expectation
value of $X$ in the vacuum state is zero, and $E(\lambda)$ is 
analytic in the neighborhood of $\lambda=0$.  Expanding $E(\lambda)$ to the
second order, we have
\begin{equation}
    E(\lambda) = E(0) + {1\over 2}\alpha \lambda^2 + ... \ , 
\end{equation}
where the coefficient $\alpha$ is real. It is easy to see that 
the real part of $E(\lambda)$ has a saddle point at $\lambda=0$: If
$\alpha>0$ (``diamagnetic" case), Re$E(\lambda)$ increases along the 
real axis of $\lambda$ (the ridge) and decreases along the imaginary
axis (the valley) at the same rate. On the other hand, when $\alpha<0$
(``paramagnetic" case), Re$E(\lambda)$ has a similar shape 
except that the real and imaginary axes are interchanged, i.e., the energy
surface is rotated by $90^\circ$.

The ``paramagnetic"  or ``diamagnetic" character of an observable  
$X$ can be identified from a property of the perturbation under
Wick rotation. If the resulting Euclidean action is
\begin{equation}
               S_E(\lambda) = S_E - \lambda XV_4 \ ,
\end{equation}
where $X$ is real function of the fields and $V_4$ is a four-volume, 
the observable is paramagnetic. This is easily seen from the
second-order derivative of the vacuum energy at $\lambda=0$, 
\begin{equation}
{d^2 E(\lambda)\over d\lambda^2} = - \int [D\phi] X^2 \exp(-S_E)\left/\int [D\phi]
        \exp(-S_E) \right. \ , 
\end{equation}
which is negative definite because $X^2$ and the integration measure 
are both positive. On the other hand, if the Euclidean action is,
\begin{equation}
            S_E(\lambda) = S_E - i\lambda XV_4 \ ,
\end{equation}
(with an extra factor of $i$), the observable is diamagnetic.  
A general observable can be a mixture of two kinds. 
In QCD, the topological charge density $F\tilde F$ is 
diamagnetic, whereas the scalar density $\overline{\psi}\psi$ is paramagnetic.
 
For any finite four-volume $V_4$, there is no spontaneous symmetry 
breaking, and the energy response remains a saddle. How
does the saddle change in the spontaneously symmetry broken 
phase as $V_4$ goes to infinity? The answer to this question 
is crucial to finding a diagnosis for possible spontaneous 
symmetry breaking.

For definiteness, we focus first on the ``paramagnetic" case. 
When spontaneous symmetry breaking occurs, the system develops 
multiple vacua (an infinite number in the case of a continuous 
symmetry breaking). Since the vacuum energy is the same in all the vacua, the
symmetry breaking perturbation $\lambda XV_4$ selects a 
particular vacuum as the physical vacuum in the path integral.  For real
$\lambda$, the physical vacuum is determined by minimization of 
the first-order energy density perturbation
\begin{equation}
        \lambda \langle {\rm vac} |X|{\rm vac} \rangle  \ . 
\end{equation}
The result, of course, depends on the sign of $\lambda$. Let us use 
$|0\rangle$ to denote the physical vacuum when $\lambda> 0$ and
$|0'\rangle$ to denote $\lambda< 0$. It is well known that $E(\lambda)$ 
has a cusp at $\lambda=0$ and $V_4=\infty$.  We will soon discuss
the physical mechanism of developing this cusp as $V_4$ 
approaches the infinity.
 
The energy density in the symmetry breaking phase can be continued 
off the real axis into the complex $\lambda$ plane.  When Re$\lambda>
0$, the path integral again selects $|0\rangle$ as the unperturbed 
ground state. We use $E^+(\lambda)$ to denote this branch of the
vacuum energy.  Likewise, when Re$\lambda< 0$, the path integral 
selects $|0'\rangle$ as the vacuum, and the corresponding energy is
denoted by $E^-(\lambda)$.

The interesting physics happens on the imaginary axis of $\lambda$. 
To see this, we start with a finite volume $V_4$ at which $E(\lambda,
V_4)$ is an analytic function of $\lambda$ in the neighborhood of 
$\lambda=0$. Slightly away from the origin and along the imaginary
axis, there are isolated zeros of the partition function: 
Since $\lambda$ is pure imaginary, there is no clear preference in the path
integral for any particular vacuum state. The contributions from 
different vacua have different phase factors.  The partition function
zeros result from cancellation among contributions of different 
vacua. This can be clearly seen in a simple example discussed in Ref.
\cite{azcoiti}.  A zero of the partition function generates a logarithmic 
cut in the free energy, and we will choose the cut along the
imaginary axis that goes to $\pm i\infty$, depending on whether 
the zero is in the upper or lower half of the plane. These logarithmic
singularities are the seeds for developing the cusp in $E({\rm Re}\lambda)$ 
at $\lambda=0$ and $V_4=\infty$. Indeed, as $V_4$ becomes
large, the phase factor is a rapid oscillating function of $\lambda$, 
and the density of the singularities grow. Meanwhile, these
singularities march towards the origin from both sides of the real 
axis and finally pinch the $\lambda=0$ point at $V_4=\infty$. Along
the imaginary axis, all logarithmic cuts coalesce into a single cut. 
This picture of cusp generation is analogous to Lee and Yang's theory
of phase transitions in a gaseous system \cite{leeyang}.

According to the above the two branches of the vacuum energy 
from different sides of the imaginary axis do not coincide on the axis. In
particular, the imaginary parts have a non-vanishing discontinuity 
across the cut, ${\rm Im}E^+(i{\rm Im}\lambda) \ne{\rm Im}E^-(i{\rm
Im}\lambda)$.  On the other hand, the real parts are equal across 
the logarithmic cut at any finite volume, $\lim_{\epsilon\rightarrow 0}
{\rm Re}E(i{\rm Im}\lambda-\epsilon, V_4)
 = {\rm Re}E(i{\rm Im}\lambda+\epsilon, V_4)$).  
The equality shall survive the infinite volume limit, and the
the result can be defined as {\it the} vacuum energy at imaginary $\lambda$.  [In
the example presented in Ref. \cite{azcoiti}, the 
infinite volume limit 
seems undefined. In a real situation, the volume cannot be
measured with an infinite accuracy, and the energy must be averaged 
over a range of volume. The oscillatory term is hence eliminated.] As
a consequence, Re$E(\lambda)$ is still a continuous function in the 
neighborhood of $\lambda=0$, but the first order derivative of $E$ is
now discontinuous everywhere across the imaginary $\lambda$, i.e., the 
cusp at $\lambda=0$ is extended to the entire neighborhood on the
imaginary axis.

The fact that Re$E^+(i{\rm Im}\lambda)$ is the same as 
Re$E^-(i{\rm Im}\lambda)$ implies that the vacuum energy has no sensitivity to
spontaneous symmetry breaking on the imaginary axis of $\lambda$, 
and in particular, it cannot be used as an observable to diagnose the
presence of such symmetry breaking.  As we have remarked earlier, 
at an imaginary $\lambda$ the symmetry breaking term $\lambda XV_4$
does not serve to select a particular vacuum state in the path integral, 
and the vacuum energy calculated can be thought of as averaged
over all the vacua. As such, it has no linear dependence in 
$\lambda$ even in the symmetry breaking phase,
\begin{equation}
    {\rm Re}E(i{\rm Im}\lambda)=E -{1\over 2}\alpha({\rm Im}\lambda)^2 \ . 
\end{equation}
This form is exactly the same as that in the symmetric phase 
considered earlier. The expectation value of $X$
calculated using this energy response vanishes, corresponding to 
the result averaged over all vacua.

The change of the vacuum energy response from the symmetric 
(a saddle) to the symmetry-breaking phase happens as follows: 
in the neighborhood of $\lambda=0$ and along the ridge, the 
energy is still a quadratic (with a positive curvature) function 
of $\lambda$. In the direction 
of valley, however, it is a linear function of $\lambda$ at each side 
of the ridge. Because the slopes at both sides are different, the vacuum
energy has a cusp at every point along the ridge. 

We are ready to consider the ``diamagnetic" case relevant to Vafa and 
Witten's proposition. Here the vacuum energy response is similar to the
``paramagnetic" case, except the real and imaginary axes are interchanged.  
The energy along the real axis is now generically a growing
quadratic function of the external field $\lambda$ as in the
symmetric phase. As such, in vector-like gauge theories,
the energy perturbation generated by pure gluonic parity-breaking 
observables {\it always increases} away from $\lambda=0$, {\it independent
of the mode} of symmetry realization. Vafa and Witten
used this growth to conclude that parity cannot be spontaneously
broken in QCD \cite{vafawitten}. However, the above 
discussion shows that even if spontaneous parity 
breaking does occur, the energy as a function of real 
$\lambda$ has the same growing behavior as in the unbroken phase.

How does one find the trace of spontaneous symmetry breaking in the 
vacuum energy when an diamagnetic observable is perturbed?
According to the above, the answer is simply to rotate the external field 
to the imaginary axis. The coefficient of the linear term in $\lambda$ is
the expectation value of the observable in the symmetry-breaking vacuum.
Note that for a general observable which is a mixture of paramagnetic
and diamagnetic characters, the special direction along which the 
vacuum energy response is insensitive to spontaneous symmetry breaking
is a straight line in the complex plane passing through the origin.

A proof that the parity-odd condensate 
$\langle 0| \overline \psi i\gamma_5\psi|0\rangle$ vanishes 
in massive QCD can be found in Ref. \cite{bitar}. This may be 
sufficient to conclude that no spontaneous 
parity breaking occurs for the real world QCD. 
For massless QCD, it is possible to direct the chiral 
condensate to the $\overline{\psi}i\gamma_5 \tau_3\psi$
direction by applying an appropriate external 
field \cite{bitar,sharpe}. Althought one can make a 
chiral transformation to rotate the condensate back 
to the $\overline{\psi}\psi$ direction and hence a new form of 
parity remains conserved, parity breaking effects 
may be seen by coupling the massless QCD to electromagnetism.

In summary, we have shown there exists a special direction in the vacuum 
energy response surface on the complex plane of probing fields
along which spontaneous symmetry breaking causes no special effect.  
This direction is just the real axis for diamagnetic observables 
in vector-like gauge theories. Therefore, although there is little doubt
that spontaneous parity breaking does not occur in massive QCD, the 
Vafa-Witten proof, however, is invalid. 

I thank J. W. Chen, T. Cohen, and P. Hoodbhoy for discussions. This work
was supported by the U. S. Department of Energy via 
grant DE-FG02-93ER-40762.

\end{document}